\documentclass[multphys,vecphys]{svmult}

\usepackage{makeidx}     
\usepackage{graphicx}    
\usepackage{multicol}    

\makeindex             

\def\beq{\begin{equation}}
\def\eeq{\end{equation}}

\begin{document}

\title*{Asteroids. From Observations to Models}
\author{D. Hestroffer\inst{1}
       \and
      P. Tanga\inst{2}}
\institute{ 
IMCCE, UMR CNRS 8028, Observatoire de Paris. 
77 Av. Denfert Rochereau F-75014 Paris, France
\texttt{hestro@imcce.fr}
\and 
Observatoire de Nice, D\'ept. Cassini, B.P. 4229, F-06304 Nice, France
\texttt{tanga@obs-nice.fr} 
}

\maketitle

\abstract Abstract
We will discuss some specific applications to the rotation state and the shapes of moderately large asteroids, and techniques of observations 
putting some emphasis on the HST/FGS instrument.

\section{Introduction}
\label{S:intro}
Although their name suggest their are point-like, asteroids are from long time well known 
to show variations in their lightcurves with shape and rotation \cite{russel06}. Observed lightcurves can for instance be explained by spinning 
tri-axial ellipsoids, but even better by convex shapes in uniform rotation \cite{magnusson89, kaasalainen02b}. Also the rotation period of these bodies seem to have some connection with their size. For instance asteroids larger than approximately 0.15\,km do not spin faster than $\approx 10\,$cycles/day (see Fig.~\ref{F:pravec_rot}). A. Harris \cite{harris96} has suggested that this limit is not the tail of some statistical distribution but does
correspond to the limit of disruption of a gravitationally bound and cohesionless {\sl rubble pile}. We will discuss in the following on the inversion of asteroids lightcurves, on rotational state of asteroids in general, on possible figures of equilibrium for rubble pile asteroids, and illustrate how we can derive information on asteroids shape and size from high resolution observation with the HST/FGS interferometer.

\section{Lightcurves}
\label{S:lc}
\begin{figure}
\centering
\includegraphics[width=8cm]{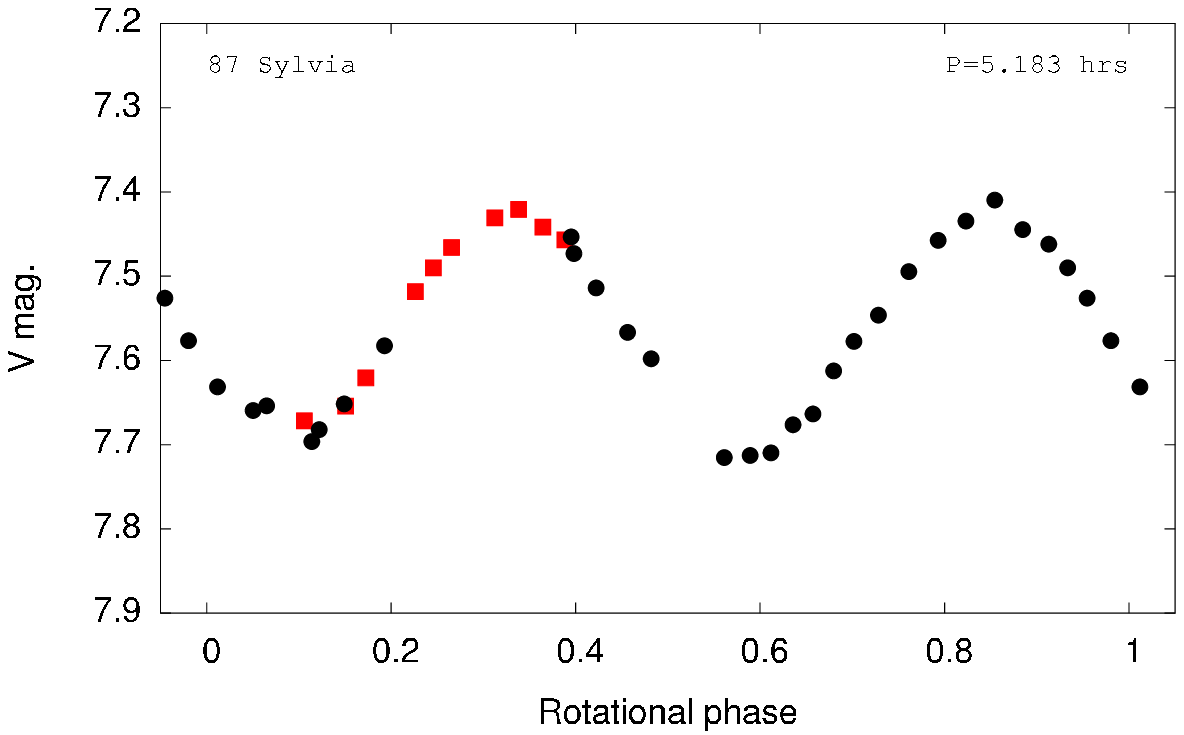}
\includegraphics[width=8cm]{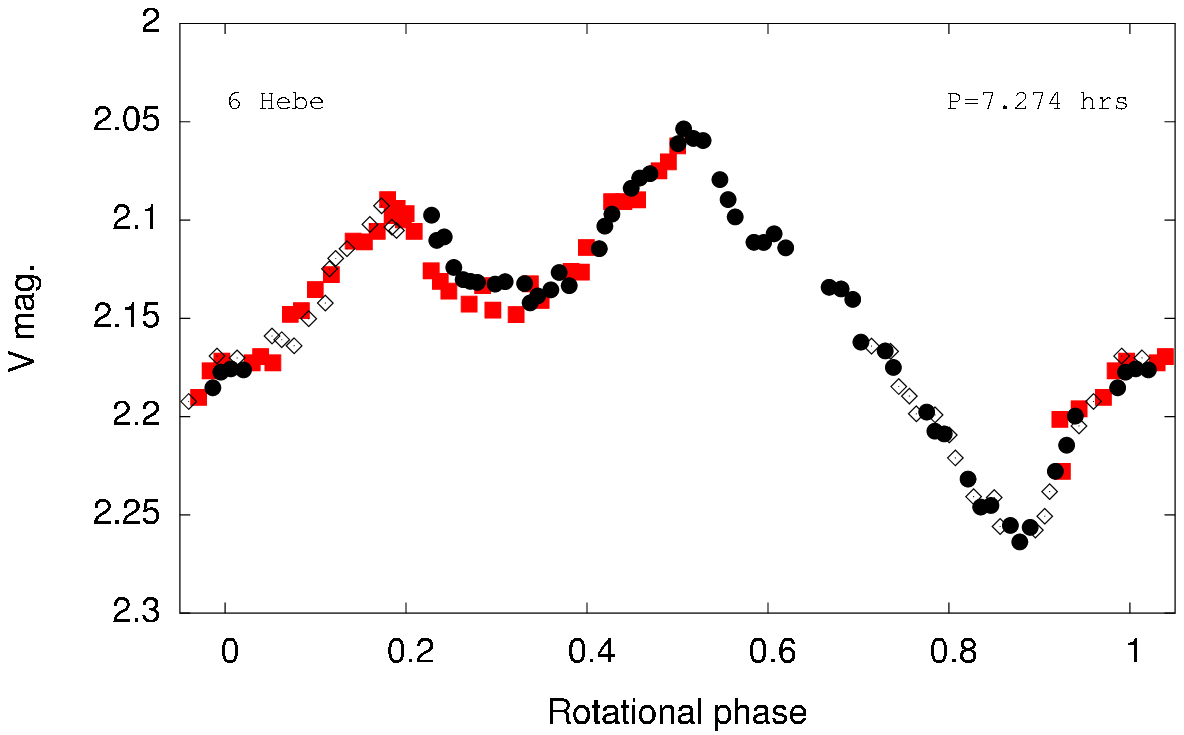}
\caption{Composite lightcurves of two asteroids from the Asteroid Photometric Catalogue \cite{magnusson89}. 
   Upper panel: (87)~Sylvia observed 
   on 1987 Feb.~3.3 ({\it filled circles}), and Feb.~6.3 ({\it filled squares}), by Weidenschilling et al. (1990). 
   Lower panel: (6)~Hebe observed on 1987 Jun. 18 ({\it filled squares}), Jun. 23 ({\it filled circles}), and Jun. 27 ({\it open squares}) by Hutton \& Blain (1988).}
\label{F:Sylvia_LC}
\end{figure}

\index{Asteroid, lightcurve}
One of the first and most important source of knowledge on the asteroids physical properties was obtained from photometric lightcurve observations. These reveal the asteroid's rotation period, their brightness variation 
due to their non spherical shape, albedo spots, light scattering of their surface, ... Two typical examples of asteroids lightcurve are given in Fig.~\ref{F:Sylvia_LC}. In particular one can see from Sylvia's composite 
lightcurve that the brightness variation is periodic, and that there are two -- almost identical -- maxima and minima. Thus Sylvia's brightness variation can be well be approached by a tri-axial ellipsoid in rotation. So that analysis of several lightcurves obtained at different apparitions provide the pole orientation and the axis ratio of the ellipsoid shape model. In the more general case however, lightcurves are not always so smoothly sinusoidal, but more irregular and asymmetric with additional extrema, ... (see the lightcurve of Hebe in Fig.~\ref{F:Sylvia_LC} for an illustration). Promising results in the lightcurve inversion problem have been obtained recently \cite{kaasalainen02c}. 

\section{Rotation}
\label{S:spin}
\index{Asteroid, rotation}
Considering a freely rotating rigid body (Euler's spinning top), integration of the Euler 
equations yields the orientation in the frame of the body of the instantaneous velocity 
\vec{\Omega}, and the orientation -- via the Euler angles $(\phi,\,\theta,\,\psi)$ -- 
of the body in a inertial frame \cite{tokiedaLNP}. As noted  
in~\cite{landau60} such a dynamical system has two -- generally non commensurable -- 
frequencies, so that the body never shows the same aspect in time. Although this exists in the solar system, there are however only a very 
few small bodies that are known to clearly show (or suspected to show) such a spin state \cite{paolicchi02}: among them one comet 
P/Halley \cite{samarasinha91}, and one asteroid asteroid (4179)~Toutatis \cite{ostro99b}. Moreover all these objects 
have relatively long rotation (spin) period. We shall see later the reason for this lack of (fast) precessing bodies.

\begin{figure}
\centering
\includegraphics[width=7cm]{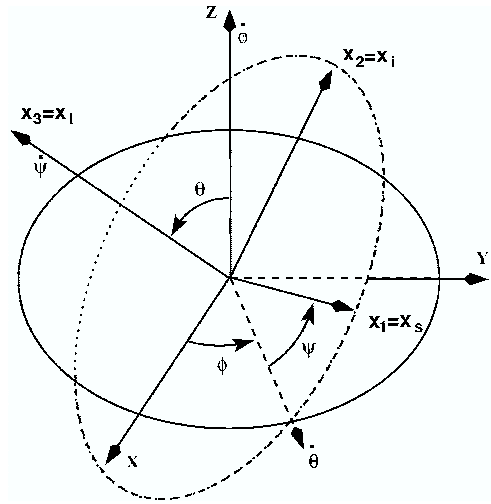}
\caption{Euler angles $(\phi,\theta,\psi)$ of the body frame $(x_{1},x_{2},x_{3})$ given in an inertial reference frame $(x,y,z)$.}
\label{F:fig_euler}
\end{figure}

\begin{figure}
\centering
\includegraphics[width=10cm]{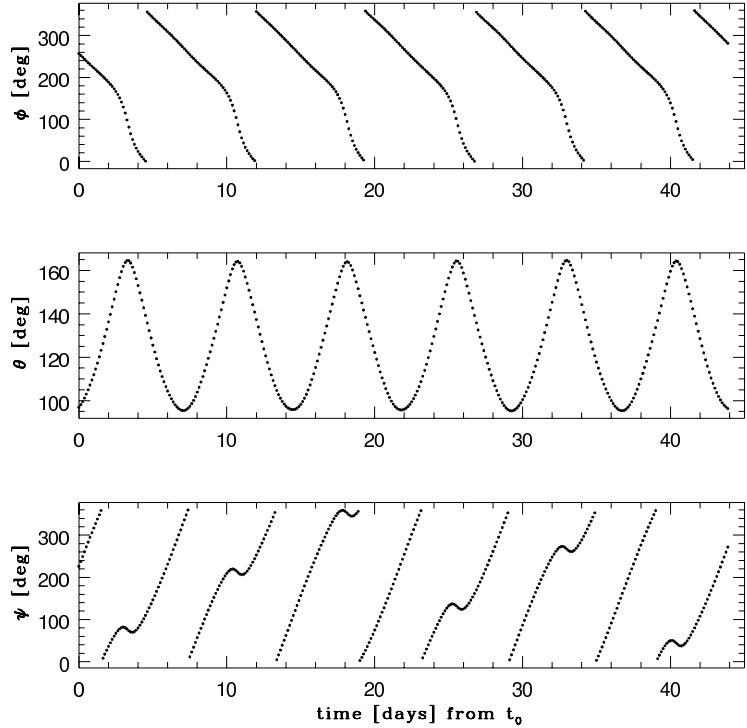}
\caption{Euler angles for (4179) Toutatis. Note the non uniformity of the circulation for angles $\phi$ and $\psi$.}
\label{F:euler}
\end{figure}

Let us remind that integration of the Euler equations:
\begin{eqnarray}
\label{E:omega}
I_1 \,\dot\omega_1 +(I_3-I_2)\,\omega_2\,\omega_3 &=& 0 \cr\cr
I_2 \,\dot\omega_2 +(I_1-I_3)\,\omega_3\,\omega_1 &=& 0 \cr\cr
I_3 \,\dot\omega_3 +(I_2-I_1)\,\omega_1\,\omega_2 &=& 0 
\end{eqnarray}
and
\begin{eqnarray}
\label{E:euler}
\dot\phi   &=& {\omega_1\sin\psi+\omega_2\cos\psi \over \sin\theta} \cr\cr
\dot\psi   &=& \omega_3 - \cos\theta\,\dot\phi= \omega_3 - {\cos\theta\,(\omega_1\sin\psi+\omega_2\cos\psi) \over
\sin\theta} \cr\cr
\dot\theta &=& \omega_1\cos\psi-\omega_2\sin\psi
\end{eqnarray}
gives the orientation of the asteroid principal axes with respect to an inertial frame 
at any epoch $t$ \cite{tokiedaLNP,landau60}. In a previous chapter by Tokieda we have also seen 
that the conservation of kinetic energy and angular momentum provides two integral relations:
\begin{eqnarray}
 2\,E &=& I_1\,\omega_1^2+I_2\,\omega_2^2+I_3\,\omega_3^2 \cr\cr
  M^2 &=& I_1^2\,\omega_1^2+I_2^2\,\omega_2^2+I_3^2\,\omega_3^2
\end{eqnarray}
The $(x_1,x_2,x_3)$ body-frame (see Fig.~\ref{F:fig_euler}) is a 
right-handed frame associated to $(x_s,x_i,x_l)$, and with this choice of indexing we have put
$I_1>I_2>I_3$. In general the Euler angles are given with respect to 
a inertial frame where the $z-$axis is -- for commodity -- aligned with the angular moment, and are understood as rotation, precession and nutation angles. In that case both angles $\theta$ and $\psi$ are periodic functions of commensurable period. Putting:
\[
  k^2 = {(I_2-I_1)\,(2\,E\,I_3-M^2)\over(I_3-I_2)\,(M^2-2\,E\,I_1)} <1
\]
and making use of the elliptic integral of the first kind
  $K = \int_0^{\pi/2}(1-k^2\sin^2u)^{-1/2}\;du$,
the period of the rotation angle $\tilde\psi$ is \cite{landau60}:
\begin{equation}
\label{E:period}
  T_{\rm rot} = 4\,K\,\sqrt{I_1\,I_2\,I_3 \over (I_3-I_2) (M^2-2\,E\,I_1)} 
\end{equation}
while the period of nutation $\theta$ is $T_{\rm nut}=T_{\rm rot}/2$.
Last, Landau \& Lifchitz \cite{landau60} have shown that the angle $\phi$ can be obtained as a sum of two periodic functions $\phi(t)=\phi_{1}(t)+\phi_{2}/({t})$ where the period of $\phi_{1}(t)$ is exactly $T_{\rm rot}/2$ and the period of 
$\phi_{2}(t)$ is $T'$, which in the general case is not commensurable to $T_{\rm rot}$. The latter period can be obtained from $2\,\pi\,T_{\rm rot}/\int_0^{T_{\rm rot}}\dot\phi(t)\;dt$.
For celestial bodies one may prefer to express the Euler angles in the frame of the ecliptic J2000.
In that case the two fundamental frequencies are mixed, and the Euler angles can be expressed as a sum of periodic functions of non-commensurable periods. Interestingly, none of the Euler angles is varying 
uniformly with time in the general case, which gave the name 'tumbling'\index{tumbling} to the asteroid 
Toutatis spin state\footnote{See an animation on URL\\
 \hbox{\tt http://www.star.ucl.ac.uk/~apod/solarsys/cap/ast/toutspin.htm}}. In the case of Toutatis, two of the principal axes have approximately the 
same moment of inertia $(I2 \sim I3)$, so that the nutation is small $(\dot\theta \sim 0)$ 
and the precession and rotation are circulating with non uniform velocity as shown in 
Fig.~\ref{F:euler}. Note also that, considering the readily inequalities $2\,E\,I_3 < M^2 < 2\,E\,I_1$, we can distinguish long-axis mode rotation (LAM, which is the case for Toutatis\footnote{This is the reason of our choice of ordering $I_1>I_2>I_3$, when the most often used one is to put $I_3>I_2>I_1$. With the presently adopted convention we only ensure 
that the rotation is associated to the Euler angle $\psi$ and axis $x_{3}$ etc.} when $M^2 > 2\,E\,I_2$, from short-axis mode rotation (SAM) ($M^2 < 2\,E\,I_2$).
This denomination reflects the fact that the instantaneous rotation 
axis is closer to the long axis, or the short axis of the body (excluding "medium-axis rotation" that anyway are 
not stable). 

\begin{figure}
\centering
\includegraphics[width=10cm]{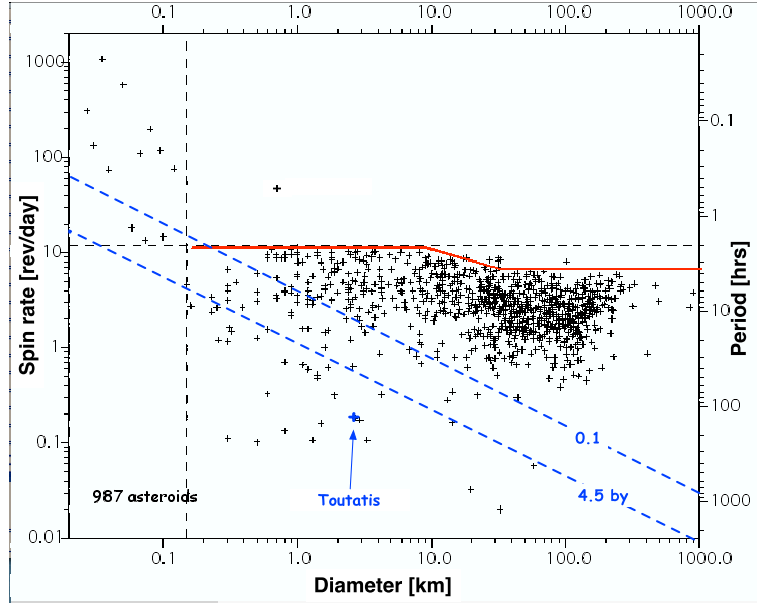}
\caption{Spin periods versus size of 987 known asteroids (from P. Pravec.) and damping time scales  ({\it dashed lines}). A limit in rotation rate for large bodies has been drawn by hand.}
\label{F:pravec_rot}
\end{figure}

One can thus expect that rigid bodies that suffer collisions and/or external perturbation can be in 
complex rotational state (misaligned rotation). 
In the general case however the asteroids do not show such large free precession/nutation tumble or wobble
over time span of typically a few days\footnote{Precession due to planetary perturbation is not discussed here.}. In contrast, it is observed that they are spinning around an axis that 1) approximately 
coincides with the direction of maximal inertia, and 2) which direction is approximately fixed in (inertial) space. 
Asteroids\index{Asteroid, damping}, as the majority of celestial objects, are not infinitely rigid bodies but are deformed under stresses. When not in principal axis spin-state, there is a loss in kinetic energy: during each wobble period a portion of the stored stress-strain energy is dissipated in the asteroid's interior. Since the angular momentum is conserved and $2T = M^2/I$ is decreasing, this dissipation is balanced by a spin state that evolves asymptotically in time toward a rotation along the axis of largest inertia ($I_1$ with the notation adopted here). The timescale of this damping process has been analyzed by \cite{burns73}. It depends on the spin rate $\Omega$, the shape $K_{1}$ and size $D$ of the body, its density $\rho$, and of course on its rigidity $\mu$:
\beq
  \tau_{\rm d} \sim \mu\,Q/(\rho\, K_1^2\, (D/2)^2\, \Omega^3)
\eeq 
where $Q$ is a quality factor expressing the ratio per wobble period of the energy contained in the oscillation to the energy lost. If the body has no rigidity it will instantaneously align its spin axis, and if it is infinitely rigid there is no evolution. Considering values typical of asteroids, \cite{harris94} alternatively gives the damping timescale by:
\[
  \tau_{\rm d} \simeq P^3/(17^3\,D^2) \quad  \hbox{\rm [in $10^9$ yr]}
\]
where $P$ is the rotation period given in hour, and $D$ is the diameter in kilometer. Asteroids densities are (roughly) in the range 1-3, their shape gives $K_{1}^2\simeq 0.01-0.1$. The value of the product $\mu\,Q$ is not known and is still in debate. Burns \& Safranov used a value of $\approx3\times10^{4}$ (cgs units), on another hand Harris \cite{harris94} -- by considering rubble-piles instead of non-porous solid rock and based on available estimates from an analysis of Phobos orbit -- adopted a value of $5\times10^{12}\,$Pa, while Efroimsky \cite{efroimsky02} states that the damping time would be 2 to 6 orders of magnitude shorter. Damping timescales with values adopted from Harris are plotted in Fig.\ref{F:pravec_rot}. 
One notes that either small asteroids and/or long period asteroids are more likely candidates for tumbling rotations -- as is the case for Toutatis -- since their damping timescale is close to the age of the Solar System. On the other hand all asteroids larger than $\approx 1\,$km in diameter and with a rotation period of less than about 10 hours should have their spin axis aligned with the largest moment of inertia axis after 100 million years. This damping hence explains that the vast majority of asteroids have single periodic lightcurves, and exhibit an invariable spin direction aligned with -- or close to -- their shortest axis of figure and also their axis of angular momentum (We do not mention long period precession/nutation which are not easily accessible with present ground-based photometry). 

\section{Figures of Equilibrium}
\label{S:3}
\index{Equilibrium, figures}
\index{Asteroid, rubble-pile}
As seen previously, we know that (disruptive) catastrophic collisions can result in a {\em rubble-pile} asteroid, i.e. gravitationally bound re-accumulated aggregates with no internal cohesion \cite{michelLNP}. It has been shown in~\cite{burns75} that asteroids could be fractured but still gravitationally bound without losing their integrity from the centrifugal forces. Here we are interested in some particular class of asteroids that excludes the few "giants" 
bodies (Ceres, Pallas, Vesta,...), and the smallest ones which are supposed to be fully cohesive rocks. 
The limits in size of the rubble-pile category are not well-defined but could be in the range 1km to 300km \cite{farinella82}. Pravec \& Harris \cite{pravec00}, noting that fast rotators can hardly have negligible tensile strength, claim that bodies larger than $\approx0.15\,$km in size are cohesionless rubble-piles (see Fig.~\ref{F:pravec_rot}). Also Britt et al. \cite{britt02}, from an analysis of known asteroid densities, defines two categories of shattered objects 
among such rubble-pile asteroids. These authors distinguish the fractured or heavily shattered objects with porosity in the range 10-25\% from the loosely consolidated rubble-piles with porosities in the range 30-80\%. Last, a classification that considers the relative tensile strength together with the porosity has been proposed \cite{richardson02}. 
Since such a rubble-pile 
asteroid should be a cohesionless and gravity-dominated body, one can expect it to have a 
figure of equilibrium. Isaac Newton, back in 1687 in his {\sl Principia}, derived the flattening ($\epsilon\equiv 1-{\hbox{polar radius} \over \hbox{equatorial radius}}$) of the 
Earth by considering it as a fluid of constant density and equalizing the weight of the water as due to gravitational and centrifugal acceleration in two radial 
canals, one directed toward the pole and another directed toward the equator. He could so explain from his theory of gravitation that the equator is not more submerged by the oceans than the poles. The flattening derived by Newton\footnote{Ch. Huygens in 1690 had a very similar approach but fundamentally different in that he did not believe in 
Newton's gravitation law and his calculation yield $\epsilon=1/578$, i.e. as if all the mass of the Earth were concentrated in it center.} $\epsilon=1/230$ is that of the rotating Earth considered as an incompressible fluid, yet different from the modern value $1/298.3$. 
Following the work of Newton other mathematicians studied the figures of equilibrium 
of a rotating mass, the reader is referred to one of the most comprehensive work that was made by S.~Chandrasekhar \cite{chandra87}. 
A few decades after the work of Newton, Maclaurin (1742) in England showed that ellipsoids of revolution are equilibrium figures of homogeneous mass of fluid in rotation, and A. Clairaut (1743) in France 
gave a relation between the density distribution inside the Earth and its flattening at its 
surface. This was followed one century later by the result of Jacobi (1834), who showed that there also exist a class of tri-axial ellipsoids for the figure of hydrostatic equilibrium. 
We will see three approaches to this problem considering different rheology: incompressible fluid, compressible material in the linear-elastic regime, and plastic-elastic material before yield.

\subsection{Hydrostatic equilibrium}
\index{Hydrostatic equilibrium}
For an incompressible fluid at rest to be in equilibrium the pressure and centrifugal force 
must balance the gravity. 
The equation of hydrostatic equilibrium states that, for each element of volume, the external body force and the boundary surface force must balance; so that considering the force per unit mass $f$ and the pressure $p$ we have $\rho\,\bf f = \nabla p$. When the external force represents a scalar potential field $\bf f={\rm grad} U$ and the density is constant one finds $\rho\bf{U-p}=0$, hence under gravity only $p = {2\over3}\pi G \rho^2\,(R_0^2-r^2)$. Similarly, stating that each volume element of the fluid is at rest also
results in an isotropic stress tensor $\sigma_{ij}=-p\delta_{ij}$ where $p=-\sigma_{ii}/3$ is the 
hydrostatic or mechanic pressure: i.e. all normal (and compressive) stresses are equal and the shearing stresses are zero.

Introducing uniform rotation along one axis, the force still is obtained from a potential, so that one sees from  a $\bf \nabla p = \rho\nabla U$ that the external surface must also be an equipotential surface, with equal density and pressure. This is a necessary condition for figures of hydrostatic equilibrium; to be sufficient the total energy (gravitational, kinetic, tidal, ...) has to be minimized. 
For instance, considering a flattened sphere of eccentricity $e$ -- together with fact that total mass $M$ and angular momentum $J$ are conserved -- 
one finds the figure of equilibrium by minimizing the energy 
$E=W+T$ (sum of gravitational and kinetic energy) over the two free parameters that are e.g. the density 
$\rho$ and the ellipticity $e$:
$\partial E / \partial \rho = 0$ and $\partial E / \partial e = 0$. By introducing the (diagonal) potential-energy tensor:
\[
  {\cal U}_{ii} = \int_V\rho\,x_j {\partial U \over \partial x_i}\;dx
\] and the inertia tensor $I_{ij}=\int_V \rho\,x_i\,x_j\;dx$, Chandrasekhar derives 
the virial relation in tensor form:
\[
  {\cal U}_{11} + \Omega^2\,I_{11} = {\cal U}_{22} + \Omega^2\,I_{22} = {\cal U}_{33} 
\]
For a homogeneous ellipsoid of semiaxis $(a_1\!\ge\!a_2\!\ge\!a_3)$, spinning along its shortest axis,  the gravitational potential and kinetic energies are respectively:
\[
  W=-{3\over10}{G\,M^2\over R^3}\,\sum_{i}A_i\,a_i^2 \quad\hbox{ and }\; T={5J^2\over 2M(a_{1}^2+a_{2}^2)}
\]
where $A_1, A_2$ and $A_3$ are calculated in terms of Jacobi integrals involving the ellipsoid's ratio only (see e.g. \cite{chandra87}):
\begin{eqnarray}
  A_{i} &=& (a_2/a_1)(a_3/a_1)\,\int_{0}^\infty \bigl((a_i/a_1)^2+u\bigr)^{-1}\,\Delta^{-1}\,du\;;\quad i=[1..3] \cr\cr
  \Delta &=& \sqrt{1+u+((a_2/a_1)^2+u)+((a_3/a_1)^2+u)}
\label{E:coeff}
\end{eqnarray}
Note however that the adimensional $A_i$ coefficients are defined differently to e.g. \cite{chandra62}; here we have $\sum_{i}A_i = 2$. 

We only give a brief and general outline of the problem resolution and analysis of the configuration's dynamical and secular stability. The reader will find a very concise treatment in~\cite{chandra87}. Here we are most interested in the figures the rotating fluid can take as a the result of the hydrostatic equilibrium hypothesis. 

\begin{figure}
\centering
\includegraphics[width=8cm]{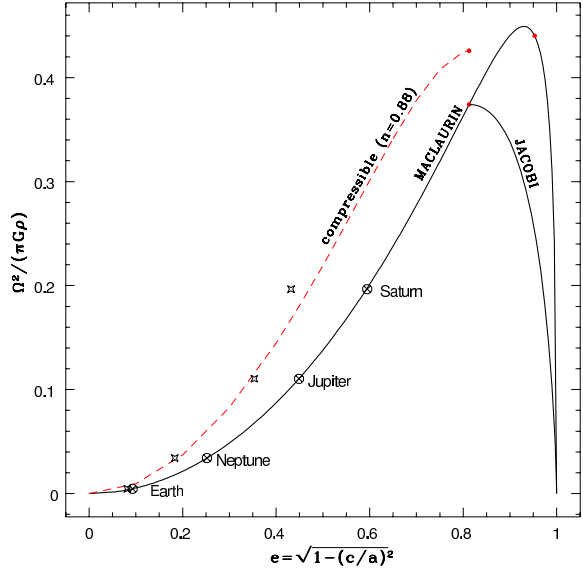}
\caption{Maclaurin and Jacobi sequences. Theoretical 
flattening of the homogeneous incompressible Earth, Jupiter, Saturn and Neptune ({\it circled crosses}). The observed 
flattening of the planets are lower ({\it crosses}). We also give 
the sequence of an compressible spheroid with a polytrope index of $n=0.88$.}
\label{F:sequenz}
\end{figure}

If there is no rotation, the equipotentials are spherical and so is the figure of equilibrium. 
A mass of homogeneous fluid with a relatively small ratio of rotational to gravitational potential 
energy $T/|W|$ will resemble a Maclaurin\index{Maclaurin} spheroid (Fig.~\ref{F:sequenz}), i.e. a axisymmetric spheroid with $a_1=a_2>a_3>0$. Then by putting the ellipticity:
\begin{equation}
  e \equiv \sqrt{1-(a_3/a_1)^2}
\end{equation}
we have:
\begin{eqnarray}
  A_1 &=& A_2\cr\cr
  A_1 &=& {1\over e^3}\,\left( \sin^{-1}e -e\sqrt{1-e^2} \right)\cr\cr
  A_3 &=& {1\over e^3}\,\left( {e\over\sqrt{1-e^2}} -\sin^{-1}e \right)
\end{eqnarray}
and the rotation frequency is given by: 
\begin{equation}
   \bar\Omega^2 \equiv {\Omega^2 \over \pi G\rho} = 2\,{\sqrt{1-e^2} \over e^2}\, \Bigl((3-2e^2)\,{\sin^{-1}e \over e} - 3\,\sqrt{1-e^2}\Bigr)
\end{equation}

By increasing for instance the rotational energy, the fluid will evolve along this sequence through flatter configurations. At a sufficiently high 
$T/|W|=0.1375$, and $\Omega^2/(\pi G\rho)=0.374$, the equilibrium is secularly unstable, and one finds that the axisymmetric configuration is no longer the lowest energy 
state available. There is another sequence consisting of tri-axial ellipsoids $(a_1>a_2>a_3)$ with some specific relation $\Phi(a_1/a_2,a_2/a_3)=0$,
the Jacobi\index{Jacobi} sequence, along which a uniformly rotating incompressible fluid will now evolve\footnote{The case of viscous or non-uniformly rotating fluid, that can evolve along different ellipsoid sequences, is not discussed here.}. Another sequence at still larger $T/|W|$ that bifurcates from the Jacobi sequence at some instability 
is one yielding binary structures. The latter could explain the formation from a catastrophic collision of a binary asteroid system through rotational fission. We will focus on the result of Jacobi that is of particular interest here, since we know from asteroids lightcurves that their shapes are not ellipsoids of revolution but better approached by tri-axial ellipsoids.  
So, we can suggest that the shape of rubble-pile asteroids, as re-accumulation of a large number of aggregates with no internal cohesion, could mimic that of incompressible fluids in hydrostatic equilibrium and, depending on their angular momentum, could be Jacobi ellipsoid.

\index{Asteroid, porosity}
One possible application of such result is that knowing the shape and rotation period 
of an asteroid as derived from the lightcurve analysis, one can determine -- in the 
framework of an incompressible fluid -- its bulk density\footnote{Newton originally used this scheme 
in his {\sl Principae} to derive the density of Jupiter}. This of course assumes  that the 
asteroid shape is close to an ellipsoid and that its density is constant. We will give two 
illustrative examples. Let us first remind the relation on the shape of a Jacobi ellipsoid:
\[
  a_1^2\,a_2^2\,{A_1-A_2\over a_2^2-a_1^2} = a_3^2\,A_3
\]
or equivalently
\[
  A_1 - (a_3/a_1)^2\,A_3 = A_2 - (a_3/a_2)^2\,A_3
\]

Numerical results for Jacobi figures are given in 
~\ref{T:jacobi} together with the corresponding rotational frequency sequence:
\[
   \bar\Omega^2 = {\Omega^2 \over \pi G\rho}  = 2\,\left(A_2-(a_3/a_2)^2\,A_3\right)
\]

\begin{table}
\caption{Jacobi figures of hydrostatic equilibrium.}
\label{T:jacobi}
\begin{center}
\begin{tabular}{cccr|rcccr|rccc}
\hline
\noalign{\smallskip}
$a_2/a_1$ & $a_3/a_2$ & $\bar\Omega$ &&& $a_2/a_1$ & $a_3/a_2$ & $\bar\Omega$  &&& $a_2/a_1$ & $a_3/a_2$ & $\bar\Omega$ \cr
\noalign{\smallskip}
\hline
1.00& 0.583& 0.3742 &&&    0.65& 0.703& 0.3475 &&&    0.30& 0.865& 0.2149 \cr
0.95& 0.598& 0.3738 &&&    0.60& 0.723& 0.3373 &&&    0.25& 0.891& 0.1813 \cr
0.90& 0.613& 0.3726 &&&    0.55& 0.744& 0.3248 &&&    0.20& 0.918& 0.1436 \cr
0.85& 0.630& 0.3703 &&&    0.50& 0.767& 0.3096 &&&    0.15& 0.944& 0.1027 \cr
0.80& 0.647& 0.3668 &&&    0.45& 0.790& 0.2913 &&&    0.10& 0.968& 0.0605 \cr
0.75& 0.664& 0.3620 &&&    0.40& 0.814& 0.2696 &&&    0.05& 0.989& 0.0219 \cr
0.70& 0.683& 0.3557 &&&    0.35& 0.839& 0.2443 &&&    0.00& $\rightarrow$1& $\rightarrow$0 \cr
\hline
\end{tabular}
\end{center}
\end{table}

\begin{figure}
\centering
\includegraphics[width=9cm]{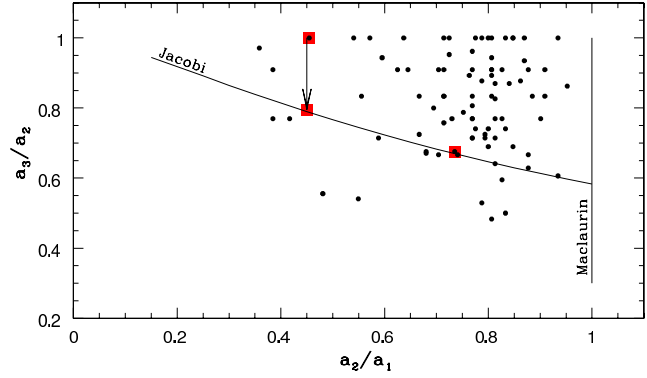}
\caption{Observed asteroids axis ratio against Maclaurin spheroids and Jacobi ellipsoids. Are overplotted the data for Eugenia on the right ({\it filled squares}), and on the left the data for the prolate spheroid Ausonia as well as the hypothetical Jacobi shape Ausonia.}
\label{F:allshap}
\end{figure}

The first example is 45 Eugenia, which body appears to have approximately the shape of a Jacobi ellipsoid 
$(a_1/a_2\sim1.35 ;\, a_2/a_3\sim1.5)$. Knowing that 
its rotation period is $P=5.7\,$hr we find a bulk density of $\rho_{\rm b} = 1.24$. Further, 
assuming that this S-type asteroid is constituted with material of density 
$\rho_{\rm g} \simeq 2.3-3$, we find a macro-porosity of $\sim45-60\,\%$. Such porosity also seems in 
good agreement with our rubble-pile hypothesis, and would roughly correspond to a random packing of aggregates. 
Moreover Eugenia is known to possess a satellite (would it be the outcome of a catastrophic 
collision, it would also enforce the rubble-pile hypothesis), that orbits its primary in 
$\approx4.7\,$days, so that one can independently estimate the bulk density 
$\rho_{\rm b} = M/V \simeq 1.2^{^{+0.6}_{-0.2}}$ \cite{merline99}. This is in good agreement with the value obtained from the hydrostatic 
fluid model. This suggests that Eugenia is a Jacobi ellipsoid, and its overall shape -- on the macroscopic scale -- 
follows equipotential surface. Moreover Eugenia could be a homogeneous body with a constant density profile. 
The second example is 63 Ausonia which shape is well approached by an ellipsoid but neither a Maclaurin 
(oblate) spheroid nor a Jacobi ellipsoid. However, the shortest axis being not very well determined by present high resolution observations, let's 
assume that the shape of Ausonia is close to that of a Jacobi ellipsoid. Then the large flattening 
of this body $(a_1/a_2\sim2.2)$ would provide a density of $\rho_{\rm b} = 0.6$. Further, assuming this C-type asteroid is constituted of material with density $\rho_{\rm g} \simeq 2-2.5$, we find a macro-porosity of 
$\sim70-76\,$\%, which value seems rather unrealistic! On the 
other hand, the observed lightcurves amplitude of Ausonia could be obtained with a binary asteroid 
where each component, in hydrostatic equilibrium \cite{leone84}, would have a somewhat higher 
density \cite{cellino85}. We shall however see in Section~\ref{S:hst} that Ausonia actually is a
single body with a shape close to a prolate spheroid. Last, if one plots the observed asteroids 
shapes against the Maclaurin and Jacobi shapes (see Fig.~\ref{F:allshap}), one clearly sees that asteroids 
shapes generally departs from equipotential surfaces.
 
If Eugenia can be considered as a Jacobi ellipsoid, it seems nevertheless that, as seen before,
asteroids in general and rubble piles in particular do not follow such figures of equilibrium for inviscid and incompressible fluids. Interesting, instead of the Jacobi ellipsoid shapes, the prolate spheroid $(a_1>a_2=a_3)$ appear to be a more common shape among asteroids. The previous analysis can be completed, in the case of stars as well as in the case of solar system bodies, by considering -- inviscid -- compressible fluids. In this case the density and pressure are no more constant through the body, and are related by some general law $f(\rho,p,T)=0$. For instance Laplace considered the relation $dp/d\rho=h\rho$, and Roche $dp/d\rho=h\rho+h'\rho^2$, $h$ and $h'$ being constants. In a similar way one can also consider bodies of incompressible but non homogeneous fluids. In addition to the numerical experiments simulating the behavior of compressible stars, let us mention the analytical work achieved in \cite{lai93} for describing the ellipsoidal figures of equilibrium in the compressible case. The authors have considered polytropes of index $n$, $p=K\rho^\Gamma$ with $\Gamma=1+1/n$. Briefly, such fluid compressibility will change the shape of the sequence as can be seen in Fig.~\ref{F:sequenz}, but not the overall shape of the figure of equilibrium. In other words, Maclaurin spheroid and Jacobi ellipsoids are still figures of equilibrium but the rotational frequency for a given shape is different to that of the incompressible case. Moreover, as shown in Fig.~\ref{F:sequenz}, for a given shape the rotational frequency is larger when considering this particular density distribution; thus the bulk density for a given shape is smaller than it is in the incompressible case. From that it also appears that the knowledge of the rotation period and geometric flattening (or shape) is not sufficient to obtain information on the density distribution inside the body. Also such compressibility would not provide a more realistic bulk density for our Ausonia example.

\subsection{Elastostatic equilibrium and elastic-plastic theories}
Another approach of interest for solid bodies and that extends the approach of fluids is to consider elastic 
bodies\footnote{And, say, thermoreologically simple bodies}. We now consider a rheology where there is a linear relation between stress $\sigma_{ij}$ and strain $\epsilon_{ij}$, and the law of constraint is given by Hooke's law \index{Hooke}:
\[
 \sigma_{ij} = \lambda\,\epsilon_{kk}\,\delta_{ij}+2\mu\,\epsilon_{ij}
\]
where $(\lambda,\mu)$ are the two Lam\'e coefficients of elasticity. The strain is derived from the deformation field $u_i(x_i)$ in the frame $(\bf x_1,x_2,x_3)$ by:
\[
  \epsilon_{ij} = {1\over2}\,\left( {\partial u_i\over\partial x_i} + {\partial u_j\over\partial x_j}\right)
\]
and the srain-stress relation can also be written as (see e.g. \cite{germain94}): 
\begin{equation}
  E\,\epsilon_{ij} = (1+\nu)\, \sigma_{ij} - \nu {\textstyle\sum_{k}}\sigma_{kk} \delta_{ij} 
\end{equation}
where $E>0$ is Young's modulus, and $-1<\nu<1/2$ the Poisson\footnote{In practice we have $\nu>0$ although negative Poisson's ratio have been witnessed in some foam \cite{lakes87},
see URL {\tt\scriptsize http://silver.neep.wisc.edu/\~{ }lakes/Poisson.html}} ratio, coefficients that only depend on the considered material. For instance, in the case of a simple compression $\sigma_{11}<0, \sigma_{22}=\sigma_{33}=0$ along axis $\bf x_1$, one has $\epsilon_{11}=1/E\,\sigma_{11}$, and the corresponding deformations rate along the perpendicular directions are $\epsilon_{22}=\epsilon_{33}=-\nu/E\,\sigma_{11} =-\nu\,\epsilon_{11}$. The couple of parameters $(E,\nu)$ is uniquely connected to the two Lam\'e coefficients. These coefficients also express the mechanic dissipation inside the material. Considering by continuity the limiting case where $\nu\rightarrow1/2$, the shearing-stiffness modulus is $\mu=0$ and one should find the results for the incompressible and inviscid fluids case. 
The static equilibrium is obtained when the sum of internal and external body forces vanishes $\bf div\,\tens{\Sigma}+f=0$, where $\tens{\Sigma}=[\sigma_{ij}]$ is the stress tensor. Considering the gravitational and centrifugal potentials from which the forces are derived $(\bf f=\rho\,grad(U))$, one obtains:
\begin{equation}
  \sigma_{ij,j} = {\partial\sigma_{ij} \over \partial x_{j}} = -f_{i} = \rho {\partial(U_{G}+U_{C}) \over \partial x_{i}} 
\label{E:equil}
\end{equation}
As seen above in Sect.~\ref{S:spin} we can assume that the object is spinning along its shortest axis 
with constant angular rate $\Omega$ and write the centrifugal potential 
\begin{displaymath}
  U_{C}=\Omega^2\,(x_{1}^2+x_{2}^2)/2
\end{displaymath} 
The gravitational potential is given by: 
\begin{displaymath}
  U_{G} = \pi\,G\,\rho\,a_{1}^3\,(U_0-\sum_i A_i\,x_{i}^2)
\end{displaymath}
where the coefficients have been defined in~(\ref{E:coeff}), and $U_{0}=\int_{0}^\infty \Delta^{-1}\,du$.
Solution of equilibrium can next be obtained by minimization of the elastic energy. The problem is analytically tractable because the loads are linear in the spatial coordinates and because of the symmetries in the considered figure. For instance this approach has been applied to the non-rigid spheroidal Earth \cite{love09}. The analytical treatment in the case of tri-axial ellipsoids is more cumbersome and has been treated in e.g. \cite{chree95,dobro82,washabaugh02}.
Considering an homogeneous, isotropic, linear-elastic, and slightly compressible material with $\nu=0.45;\,0.499$, Washabaug \& Scheeres \cite{washabaugh02} have shown that, at low angular momentum, ellipsoidal figures with compressive stresses at the surface exist (i.e. the presence of tensile strength is not needed) but they generally lie only in the vicinity the minimal energy state. They also showed that the elastic energy minima are weaker for compressible material, but nevertheless they occur at high angular momenta and for very elongated shapes which shapes are not observed among asteroids.

Eventually, we will discuss a more general approach that does not depend on the actual stress-strain behavior or possible residual stresses, but we will consider the {\sl limit} stresses of an elastic-plastic body. Starting from ~(\ref{E:equil}) -- and considering only ellipsoids -- the equilibrium equations and boundary conditions leaves three degree of freedom in the general solution. Thus one can consider that the material is fluid (imposing that the shear stresses are vanishing and all normal stresses are equal), or one can consider linear-elastic isotropic material by introducing an additional relation between strain and stresses from Hooke's law. These are the two options we have considered so far. 
Considering fluids or elastic deformations to model asteroids would not allow for large fails, boulders or 
craters on the surface of these small rocky bodies as can be witnessed for 
instance on the surface of the planetary satellite Phobos or asteroid (253)~Mathilde. 
It has been suggested that cohesionless bodies could maintain shapes significantly 
different from figures of hydrostatic equilibrium, so long as i) one assumes that the
rotating mass is not a fluid but behaves as a granular soil with non negligible
solid-to-solid friction, and ii) internal stresses are not high enough to crush 
individual particles \cite{hammergren99,holsapple01}. In this case the material can sustain non negligible 
tangential force before plastic flow. The Mohr-Coulomb\index{Mohr-Coulomb} criteria is generally used for deriving the maximal
stress strain before yield of a given soil. The stress tensor $\tens{\Sigma}=[\sigma_{ij}]$ being symmetric, it is diagonalizable; given the principal stresses $\sigma_{1},\,\sigma_{2},\,\sigma_{3}$ and neglecting the cohesive strength, the Mohr-Coulomb criteria states:
\[
  \tan\phi\ge {\sigma_{1}/\sigma_{3}-1 \over 2 \sqrt{\sigma_{1}/\sigma_{3}}}
\]
that is, the maximal tangential stress is limited by only the most and least (compressive) normal stresses $sigma_{1} and \sigma_{3}$. In such case we no more have an equality that provides the unique figure of equilibrium, but instead a range of possible configurations. Considering again tri-axial ellipsoids of constant density, it appears that the ratio $sigma_{1}/\sigma_{3}$ does not depend on the spatial coordinates, and one can plot the rotational frequency versus the axis ratio for a given friction angle \cite{holsapple01}. We see in Fig.~\ref{F:holsa} that the hydrostatic equilibrium is obtained for negligible friction, and for a given friction angle (e.g. $\Phi=15\,$deg) there is a zone of possible ellipsoidal figures. That is, for those tri-axial ellipsoids the gravitational and centrifugal load are small enough to be sustained by the friction and hence avoid failure. We also see that the rotation frequency is limited at the largest possible friction $\Phi=90\,$deg, in particular for a sphere one has \cite{harris96}:
\[
    \sqrt{\bar\Omega_{\rm max}}=4\pi/3
\]
but this does depend linearly on the axis ratio. Last if one plots the data from known asteroid, assuming typical densities in the range 1--3, it eventually appears that the great majority of observed shapes is consistent with such a cohesionless Mohr-Coulomb model and a friction angle of $\approx25\,$deg, which value seems realistic and typical of dry terrestrial soils (it is $\approx30$ for sand).
\begin{figure}
\centering
\includegraphics[width=9cm]{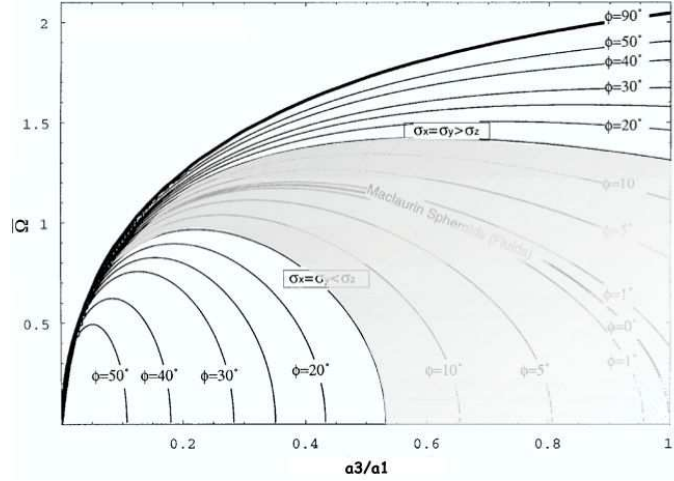}
\caption{Limit rotational frequency figures for oblate spheroids (from \cite{holsapple01}).}
\label{F:holsap}
\end{figure}

\subsection{Binary systems and the density profile}
\index{Asteroid, binary}
Asteroids do have satellites too! Since the first one that was discovered by the Galileo probe during its fly-by with Ida in 1993, about a dozen additional systems have been identified by ground-based observations (radar, adaptive optics, photometry) in the NEO or KBO population as well as in the main belt. Note that previous attempts from the Hubble Space Telescope imaging instrument were unsuccessful \cite{storrs99}. The presence of a satellite (generally 10-20 times smaller than the primary) is of high value to determine the mass of the primary from Kepler's third law and eventually its bulk density. Moreover, since these objects are rather flattened or elongated, the dynamical perturbations due to the non-spherical gravitational potential are expected to be relatively large. As seen previously the (secular) perturbations depends on the potential, or mass distribution \cite{deleflieLNP}. If the spin rate of the primary is much larger than the orbital rate of the satellite $(\Omega_{\rm rot} \gg n)$, the secular effect of the $C_{22}$ is negligible, and the major perturbation arises from the $J_2$ coefficient:
\[
  J_2=-C_{20}=-{I_1+I_2-2I_3\over 2Ma_e^2}
\]
\index{Precession, periastron}
Further, the secular terms are obtained from Lagrange equations:
\begin{eqnarray}
  \dot a &=& \dot e = \dot i=0\cr\cr
  \dot \omega &=&  {3\over4}{n\,a_e^2 \over a^2\,(1-e^2)^2}\,
		(1-5\cos^2i) \; C_{20} + ... \cr\cr
  \dot \Omega &=&  {3\over2}{n\,a_e^2 \over a^2\,(1-e^2)^2}\,
		\cos i \; C_{20} + ... \cr\cr
  \dot M &=&  {3\over4}{n\,a_e^2 \over a^2\,(1-e^2)^{3/2}}\,
		 (1-3\cos^2i) \; C_{20} + ...
\end{eqnarray}

\begin{figure}
\centering
\includegraphics[width=9cm]{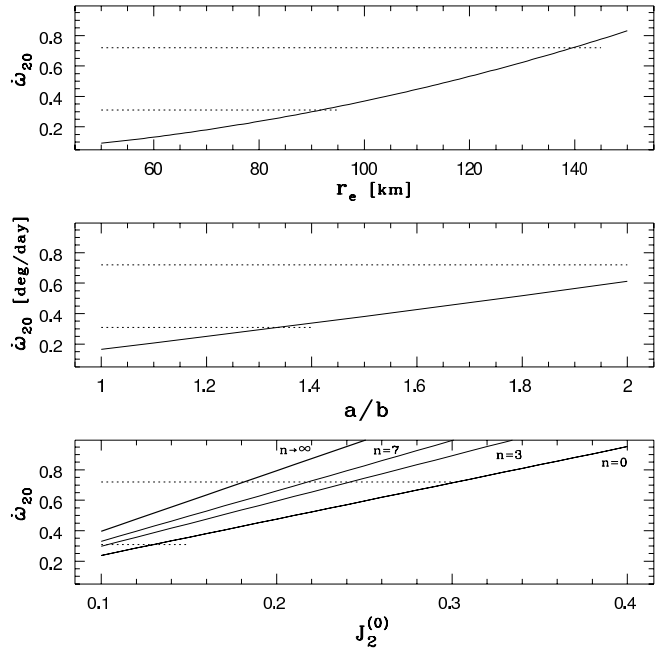}
\caption{Secular periastron advance $\dot\omega_{20}$ for the orbit of 22 Kalliope's companion.}
\label{F:kalli}
\end{figure}

We know since Clairaut and Radau that the knowledge of the dynamical and geometric flattening brings insight on the density distribution inside the body. Thus observing the secular perturbation on the satellite's orbit over successive months provides the dynamical flattening while observing photometric variation over successive apparitions provides the geometric flattening. By comparing both one can at least test the hypothesis of constant bulk-density inside the (primary) asteroid. This has been applied to the orbit of Kalliope's companion \cite{marchis03}. Assuming a homogeneous primary with constant density distribution the observed geometrical flattening provides a dynamical $J_{2}={1\over10}\,{a_1^2+ a_2^2-2\,a_3^2\over a_e^2}$ and in turn a precession rate of $\dot\omega\sim0.3\,$deg/day. This value is in severe conflict with the observed one $\approx0.7\,$deg/day. As shown in Fig.~\ref{F:kalli} there are three possible explanations that maybe all concur together to this discrepancy: the size of the body, the geometrical flattening, and the density distribution. Asteroids diameters are essentially given by an indirect method from observations with the IRAS satellite, and can be in error by 10\% or more \cite{hestro02d}. The shape being derived from disk-integrated photometric data and not high resolution imaging is not known with the best accuracy neither. Last, including an empirical but simple density distribution of the form $\rho(r,\theta,\phi)=\rho_0\,r^{\alpha};\;\alpha\in{I\kern -3pt R}$, i.e. which {\sl increases} as we progress toward the surface ($r=1$), one can write the zonal harmonic as a function of the one for $\alpha=0$, i.e. at constant density $J_{2}^{0}$:
\[
  J_2={5\over3}{\alpha+3\over \alpha+5}\,J_2^0
\]
Hence for a given shape, the pericenter precession is increased by concentrating the mass at the outer surface of the body. Assuming that the large precession is due to the non homogeneous mass distribution alone, one finds a density at Kalliope's surface of $\approx 7$, which is marginally acceptable. Let's now consider that Kalliope is a size-sorted rubble-pile, where the larger and more irregular rocks are in the central part and the smaller material is kept by friction at the outer layers \cite{britt01}. Such mass distribution could correspond to a body of homogeneous but size-sorted material with larger relative voids (or porosity) in the central part, and more densely packed material toward the surface.

\section{The determination of shape and spin parameters by Hubble Space Telescope}

\subsection{The FGS interferometer}
\index{HST/FGS}
The Hubble Space Telescope (HST) is a complex of instruments, built around a
2.4-meters telescope, orbiting the Earth. For several reasons
(pressure on observing time requests, time constraints due to the
orbit, complexity of the instruments, etc.) it is sometimes very
difficult to apply for successfully and to use.

However, its great advantage -- being outside the Earth atmosphere --
allows to obtain otherwise difficult measurements and
observations. While its imaging capabilities of deep--sky objects and
planetary surfaces/atmospheres are well known, they are not sufficient
to perform accurate measurements of asteroid shapes and sizes. In fact,
the highest resolution is reached by the Planetary Camera, having a
plate scale of 46\,mas/pixel\footnote{In the following we express all
apparent sizes in milli-arcseconds (mas) $=10^{-3}$ seconds of arc}. 
This value is of the same order than the apparent size of several,
interesting main belt objects at opposition, and allows some
resolution to be achieved on a very restricted set of the largest
bodies only.

A much more sentive instrument is the Fine Guidance Sensor (FGS), an
interferometer normally used to allow careful pointing and guiding of
the HST while imaging is performed by the main CCD cameras. Three FGS
instruments are mounted close to the focal plane, and each of them
works by producing interference between the two beams coming from the 
defocalised semi-pupils of the telescope. Inside the FGS the
beam is divided into two parts, associated to two perpendicular axis
(in the following we refer to them as the FGS-X and FGS-Y axis). Each
beam enters a Koester prism. 
Inside this device the self-interference occurs. The resulting two beams
exiting a single Koester are then collected by two photometers,
measuring their flux. The important feature to note here, is that the
difference between the fluxes is a function of the inclination of the
incoming wavefront\footnote{More details can be learned by browsing the
HST/FGS on-line handbook at URL: {\tt http://www.stsci.edu/instruments/fgs/}.}.

The FGS is able to ``scan" the focal plane in straight line, along
the FGS-X and FGS-Y axis, in steps of 1-2\,mas, and over a few
arcseconds. 
Each step corresponds to a different inclination of
the wavefront and thus to a different response of the photometers.
A response function, called ``S-curve'', is reconstructed from the
flux difference normalized to the total flux. 

\begin{figure}
\centering
\includegraphics[width=10cm]{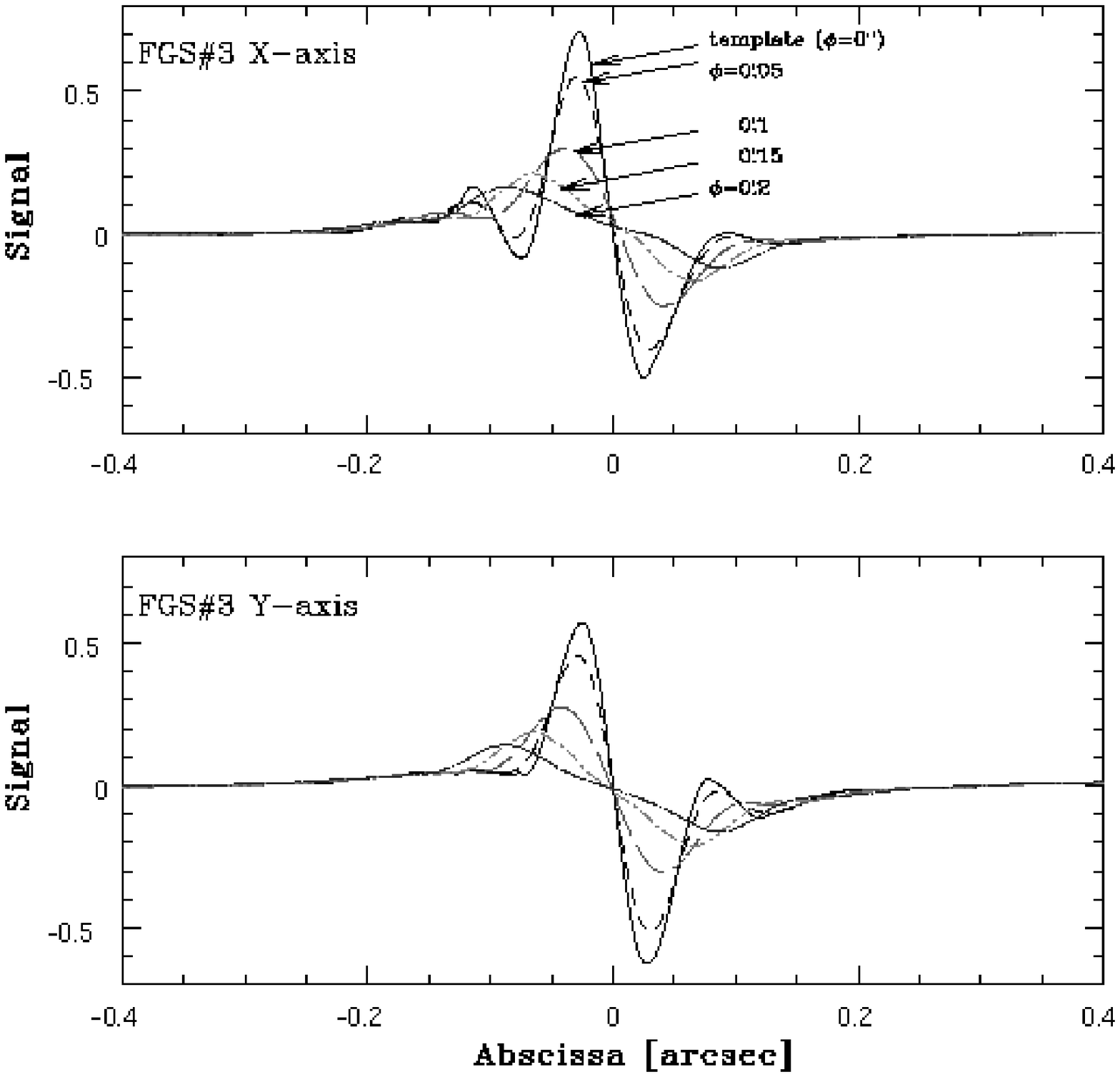}
\includegraphics[width=8cm]{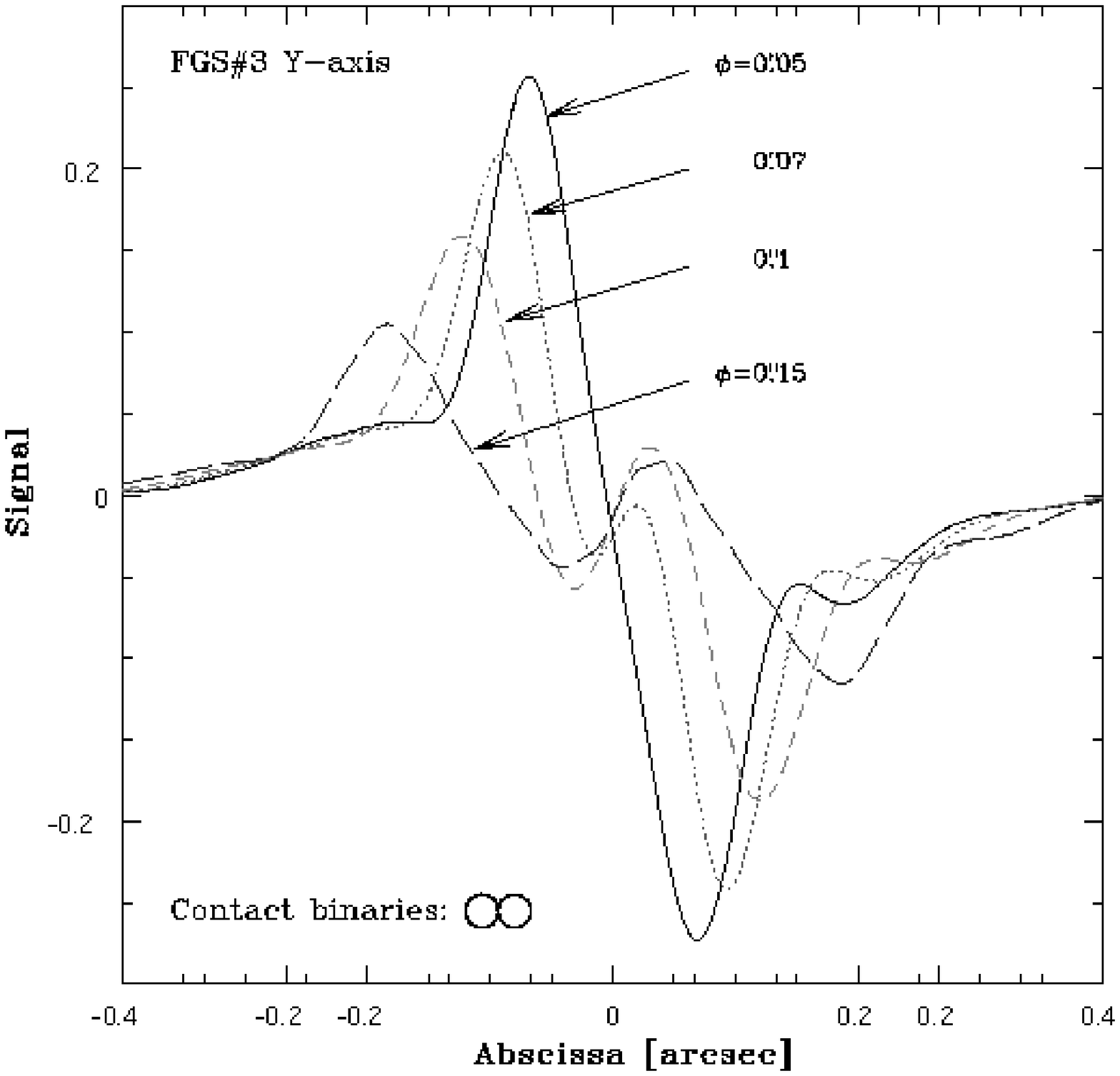}
\caption{Upper panels: the simulated response curve, along the FGS-X and FGS-Y axis,
for a uniform, circular source of different sizes (indicated in
arcsec). The ``template'' curve correspond to the observation of a
point-like star. Lower panel: the simulated response curve for a
double source composed by two equal, tangent disks of different diameters.}
\label{F:response}
\end{figure}

It is clear, then, that the response for an extended source depends
upon the light distribution on the focal plane. The FGS 
sensitivity being optimal 
for spatial frequencies $\leq 200\,$mas, it is normally employed to
measure the diameter of large star disks or the separation of close doubles.
An example of the simulated response curve for an extended, uniform disk, or for
a double disk, is given in figure (Fig. \ref{F:response}). As it can
be seen, the largest is the source, the smallest is the amplitude of
the S-curve. A double object has the tendency to double the curve
peaks.
In order to better understand the results presented in the following, 
it must be noted that each scan by the FGS produces two S-curves, one
for each axis. As a consequence, the extracted informations concern the
size of the studied object as projected along the two FGS axis.

In recent years, the FGS has been used for the first time to measure
rapidly rotating and moving objects: the asteroids.
Six objects having interesting spin and lightcurve properties have
been chosen, each observed for an HST ``orbit'', i.e. for a total
duration of about 40 minutes (corresponding to 30-40 FGS scans). 
During this time lapse two movements affect the observation: the spin of 
the object and its proper motion. The first one is highly desirable, 
since it can help to constrain the asteroid shape by studying the
variations of the FGS-X and FGS-Y projections. The second one, on the
other hand, has been minimized by choosing epochs close to the
stationary points in the asteroids apparent motion, and by applying a
correction (a posteriori) to compensate for the parallax due to HST displacement.

In summary, each set of scans for each asteroid contains informations
on its shape and size, as projected over FGS-X and FGS-Y axis, over a
limited fraction of the rotation period -- roughly corresponding, for
the given set of objects, to about the 10\% (about 30-40 degrees
of rotation). It can thus be supposed that a fitting of a model to the
set of S-curves can lead to the reconstruction of sizes and
shapes. The degree of accuracy (i.e. the number of parameters to be
determined) will depend upon the quantity of observations and their
signal/noise ratio. 

\subsection{From data to modeling}

In general, given a set of FGS-X and FGS-Y S-curve couples at each epoch, 
the fitting procedure begins by determining shape, orientation and
size of the on-sky projected ellipse that separately accommodates best to
each curve. 

The problem, then, is to identify which is the solid,
three-dimensional body that is capable of producing the observed,
projected ellipses. If a traditional equilibrium figure is searched
for, an ellipsoid with a given orientation of the spin axis will 
be the figure of choice.
 
It must be noted that traditional photometry from Earth-based
telescopes normally records brightness variations associated to the
object shape, directly yielding its rotation period. From  
lightcurves taken at different epochs, a first estimation of the spin
axis direction can be obtained. In general, however, some symmetries
in the problem does not allow to discriminate between degenerate
solutions. The result of photometric pole determination is thus
expressed by two possible spin axis directions, each yielding, for a given
epoch, the same object area projected on the sky, i.e. the same 
brightness. However, an instrument capable of directly detecting 
the orientation of the shape and its variation in time, can
immediately eliminate the ambiguity and help to discard one of 
the two solutions.
This is the first result that have been derived by HST observations. 

Thus, having selected the good pole solution between the pair available, the 
model is fitted to the S-curve is that of an ellipsoid of uniform 
brightness. The shape is described by the three ellipsoid semi-axis a, b
and c. Since the uncertainty in pole coordinates can reach
several degrees, a trial and error adjustment, reducing the O-C, is
performed. In fact, we realized that additionally solving the fit 
for pole coordinates seems not to add significant improvements 
given the set of data currently available. Thus, pole coordinates 
do not enter the core of the fit process. 

After having determined the best-fitting ellipsoid, we will have a set
of values corresponding to semi--axis sizes, spin pole coordinates and
rotational phase. In turn, this allows to re-compute for each epoch
the ellipse projected on the sky. Another iteration can thus be
performed repeating the whole process from the beginning. 
If an ellipsoidal shape solution exists, the parameters 
rapidly converges and the residuals collapse. Significant residuals
are, on the contrary, the indication of some detectable departure from
the simple assumption on the shape, and some more complex models
deserve to be taken into account.

\subsection{Some significant examples}
\label{S:hst}
In the following we illustrate some results that we think to be
particularly significant. The accuracy of the results, their
limitations and the practical difficulties should be apparent.
A complete review of results is published in \cite{hestro02I,tanga01,tanga03}, while the following table reviews the main parameters derived.
\begin{table}
\begin{center}
\caption{A summary of FGS size measurements, obtained by considering
  ellipsoidal models.
(216) Kleopatra, having a complex shape, is not shown here. 
The last column gives the ratio
of the axis; the parentheses indicate that either $b$ or $c$ are not well constrained.}
\label{T:hst_{fit}}
\renewcommand{\arraystretch}{1.2}
\begin{tabular}[t]{lrcrcc}
\hline
Name && a, b, c [km] && a/b & a/c \\
\hline
~(15) Eunomia && 181, 103, 102 && (1.76) & 1.78\\
~(43) Ariadne && 45, 26, 26 && 1.71  & (1.71)\\
~(44) Nysa    && 59, 35, 35 && (1.72) & 1.72\\
~(63) Ausonia && 75, 33, 33 && 2.28  & (2.28)\\
(624) Hektor && 62, 28, 28 && 2.21  & 2.21\\
\hline
\end{tabular}
\end{center}
\end{table}

\subsubsection{63 Ausonia}

\begin{figure}
\centering
\includegraphics[width=10cm]{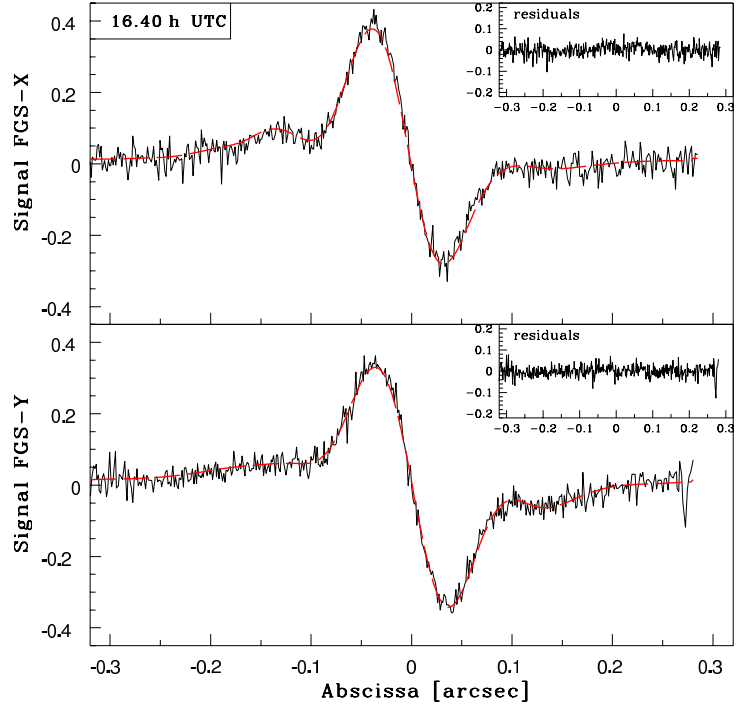}
\caption{The Ausonia S-curves at the beginning of the observation, fitted by
using the tri-axial ellipsoid given in Table~\ref{T:hst_{fit}}. The residuals of the
fit given in the inset are very small.\label{F:ausonia}}
\end{figure}

63 Ausonia was one of the brightest objects in the set, and the first
to be observed. The signal-to-noise ratio is rather good, so the
fitting process operates in ideal conditions. Figure~\ref{F:ausonia}
shows the S-curves of (63)~Ausonia for a selected epoch. 
The residuals of the fit with a
three axis ellipsoid are shown in the inset and are very small in this case, showing that 
the identified 3D ellipsoid is completely consistent with the
available FGS data, as discussed above.

Due to its orientation on the sky, the asteroid, while rotating,
exhibits some interesting variations in the length of the projected
$a$ and $b$ axis, while $c$ (coinciding with the rotation axis) does
not move and its influence on the FGS measurements is minor. For this reason, 
the value of the
{\sl c} axis is affected by a high uncertainty (around 5\,mas or more) 
while $a$ and $b$ are constrained to a 1\,mas level. These figures can
be considered to be typical for this kind of model-dependent fit. 

\subsubsection{216 Kleopatra}

This was most irregular body observed. A little time before the HST
observations, it was observed by radar \cite{ostro00}. The
reconstructed shape seemed to hint to a bi-lobated object, very
elongated and irregular.

The HST/FGS signal suggests an elongated shape, well approximated
by two - not detached - ellipsoids, whose signature is well visible in
the S-curve. The overall shape appears to be more elongated and
flattened in comparison to radar data.
\begin{figure}
\centering
\includegraphics[width=10cm]{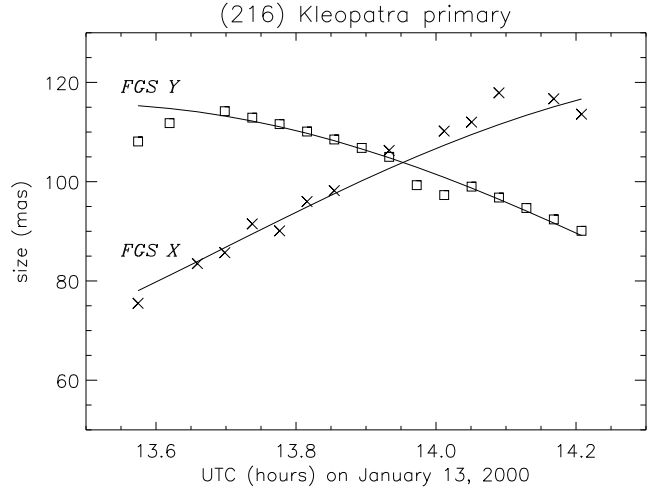}
\caption{The variation of the sizes projected on FGS-X and FGS-Y for the main
component of the double-lobed asteroid (216)~Kleopatra, during the observation.
The shape of the object and its orientation relatively to the FGS axis
are represented by the two ellipsoids below the curves.\label{kleo}}
\end{figure}
Details are given in \cite{tanga01} and \cite{hestro02b}.

\subsubsection{624 Hektor}

This Trojan asteroids is the faintest asteroid observed by HST/FGS during the
program (V=15.0 at the epoch of the observation), being at an average
distance corresponding to the semi-major axis of the orbit of
Jupiter. 
Due to its faintness, the S/N ratio is very small and it probably
represents the lowest possible for this kind of studies by the FGS. 
The response functions is best fitted by a very 
elongated shape, but due to the low S/N ratio it is not possible to
clarify if (624)~Hektor can really
be considered to be a contact binary as supposed in the past by Weidenschilling \cite{weiden80}.
\begin{figure}[tr]
\centering
\includegraphics[width=9cm]{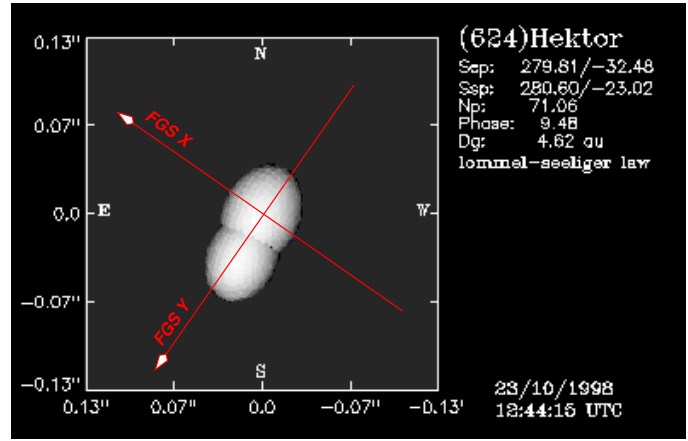}
\caption{The indicative shape of Hektor obtained by a
  best-fit of the FGS observations. The noisy data set does not allow
  to draw definitive conclusions. 
\label{F:hektor}}
\end{figure}
Figure~\ref{F:hektor} shows the best-fit shape obtained by HST
observations.\\

\subsubsection{15 Eunomia}

Together with the previous object, (15)~Eunomia is another difficult target. 
(15) Eunomia is the object having the largest apparent size, as shown
by the small amplitude of the S-curve. 
\begin{figure}[tr]
\centering
\includegraphics[width=9cm]{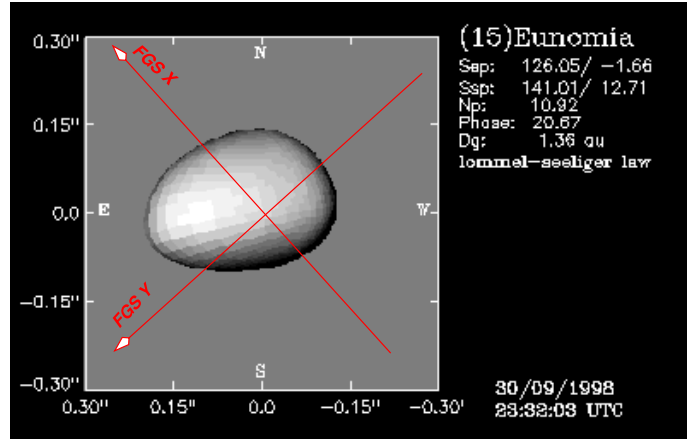}
\caption{Lower panel:
the suggested shape of (15)~Eunomia is consistent with the strongly asymmetric S-curve
obtained for this object. Other possible solutions require the use of 
other constraints, such as those coming from photometry.\label{F:eunomia}}
\end{figure}
The shape of the signal is consistent with a shape more complex than a simple ellipsoid,
as shown in Fig.~\ref{F:eunomia}. Unfortunately, in order to better
constrain its shape, more observations and a widest coverage of its
rotation would be necessary. 

\section{Conclusions}

The observations by HST/FGS are a sensitive and powerful method to
determine size and shape of asteroids. However, some limitations
should be kept in mind.

First of all, the observing time is very limited and hard to
obtain. For this reason, it is very difficult to sample the whole
rotation curve of an object. Unfortunately, this is the only way to
constrain complex shapes.

Furthermore, the shape solution that is found depends upon some
{\sl a priori} choices, such as the basic shape model, its
scattering properties, the absence of albedo markings, etc. This details, 
while not changing the qualitative interpretation of the data, are
probably important in order to define the ultimate precision of the
observations. 

Finally, the FGS can work on a limited sample of extended and bright
asteroids. However, the same basic approach, coupled with more
sophisticated data inversion techniques, can be applied to other
optical interferometers, based on the Earth surface. 
We can thus hope that, in the future, the sample of asteroids for
which size, shape and orientations are known will give us a more
complete view of their physical properties.
Last, combining photometric data to high resolution data and astrometric 
positions of asteroid satellite should bring insight of the primary asteroid interior
and possible collisional history. 

%
%
 \bibliographystyle{unsrt}
 \bibliography{astro}
%

\end{document}